# The formation conditions of enstatite chondrites: Insights from trace element geochemistry of olivine-bearing chondrules in Sahara 97096 (EH3)

Emmanuel Jacquet[1,2], Olivier Alard[3], Matthieu Gounelle[1]

[1]Institut de Minéralogie, de Physique des Matériaux et de Cosmochimie, CNRS & Muséum National d'Histoire Naturelle, UMR 7202, 57 rue Cuvier, 75005 Paris, France.

[2]Canadian Institute for Theoretical Astrophysics, 60 St George Street, Toronto, ON, M5S 3H8, Canada.

[3]Géosciences Montpellier, UMR 5243, Université de Montpellier II, Place E. Bataillon, 34095 Montpellier cedex 5, France.

E-mail: emjacquet@mnhn.fr

## *Abstract*

We report in situ LA-ICP-MS trace element analyses of silicate phases in olivine-bearing chondrules in the Sahara 97096 (EH3) enstatite chondrite. Most olivine and enstatite present rare earth element (REE) patterns comparable to their counterparts in type I chondrules in ordinary chondrites. They thus likely share a similar igneous origin, likely under similar redox conditions. The mesostasis however frequently shows negative Eu and/or Yb (and more rarely Sm) anomalies, evidently out of equilibrium with olivine and enstatite. We suggest that this reflects crystallization of oldhamite during a sulfidation event, already inferred by others, during which the mesostasis was molten, where the complementary positive Eu and Yb anomalies exhibited by oldhamite would have possibly arisen due to a divalent state of these elements. Much of this igneous oldhamite would have been expelled from the chondrules, presumably by inertial acceleration or surface tension effects, and would have contributed to the high abundance of opaque nodules found outside them in EH chondrites. In two chondrules, olivine and enstatite exhibit negatively sloped REE patterns, which may be an extreme manifestation of a general phenomenon (possibly linked to near-liquidus partitioning) underlying the overabundance of light REE observed in most chondrule silicates relative to equilibrium predictions. The silicate phases in one of these two chondrules show complementary Eu, Yb and Sm anomalies providing direct evidence for the postulated occurrence of the divalent state for these elements at some stage in the formation reservoir of enstatite chondrites. Our work supports the idea that the peculiarities of enstatite chondrites may not require a condensation sequence at high C/O ratios as has long been believed.

## *1. Introduction*

Despite their isotopic similarity to the Earth (Javoy 1995; Trinquier et al. 2007; Herwartz et al. 2014), enstatite chondrites arguably constitute the most outlandish class of chondrites (Prior 1920; Keil 1968; Weisberg & Kimura 2012). Their mineralogy is very reduced, with nearly FeO-free enstatite as the dominant silicate, common free silica; Si-bearing kamacite and, in addition to troilite, diverse sulfides largely unique to this clan such as oldhamite (CaS), niningerite (MgS) or daubréelite ($FeCr_2S_4$) ; and nitrides such as osbornite (TiN) etc. (Weisberg and Kimura 2012). Save for altered clasts in the Kaidun breccia, their fine-grained matrix seems devoid of any evidence of aqueous alteration (Rubin et al. 2009; Weisberg et al. 2014; Lehner et al. 2014).

This reduced mineralogy has long suggested that enstatite chondrites resulted from a condensation sequence in conditions more reducing than canonical. Specifically, thermodynamic calculations by Lodders & Fegley (1993) suggested that for C/O > 1, early-condensed oldhamite would have concentrated rare earth elements (REE) explaining their observed 1-2 order-of-magnitude enrichments above CI (Crozaz & Lundberg 1995; Gannoun et al. 2011). From an astrophysical standpoint, high C/O could come about in the inner disk through locking of water beyond the snow line by efficient accretion there—the "cold finger effect" of Stevenson and Lunine (1988) (see also Ciesla & Cuzzi 2006); or, alternatively, by inward drift of C-enriched dust (see e.g. Ebel and Alexander 2011). However, notwithstanding the minute amounts of C-bearing presolar grains, evidence for expected collateral condensates such as silicon or titanium carbides is missing, and the few calcium-aluminum-rich inclusions (CAIs) described in enstatite chondrites are quite similar to their counterparts in other chondrite groups (Weisberg and Kimura 2012).

Moreover, enstatite chondrites preserve a record of a stage of more oxidized conditions. FeO-bearing silicates, having undergone varying degrees of reduction, have been repeatedly described in the unequilibrated enstatite chondrites (Rambaldi et al. 1983; Lusby et al. 1987; Weisberg et al. 1994; Kimura et al. 2003; Weisberg et al. 2011), and appear to belong to the same oxygen isotopic reservoir as the host chondrite (Kimura et al. 2003; Weisberg et al. 2011). Based on physico-chemical investigations of niningerite-troilite-silica-bearing chondrules in Sahara 97072, Lehner et al. (2013) proposed that the enstatite chondrite sulfides actually result from a sulfidation event on originally more ferroan silicates (see also Piani et al. 2013; Lehner et al. 2014).

Among these witnesses from putative earlier formation conditions, olivine is of especial interest (Binns 1967; Rambaldi et al 1983; Ikeda 1989a; Weisberg et al. 2011). Indeed while porphyritic olivine or olivine-pyroxene chondrules make up only about 4 % of chondrules in unequilibrated enstatite chondrites (Jones 2012), olivine does dominate the mineralogy of other chondrite groups, in particular their chondrules. This is consistent with olivine being predicted to be a major product of equilibrium condensation from a gas of solar conditions, which, mutatis mutandis, is expected to have given birth to chondrule precursors. So if the peculiarities of enstatite chondrites are not due to a condensation sequence under high C/O ratios, olivine is likely to have been a major constituent of precursors of enstatite chondrite components at some point in the "prehistory" of these meteorites. Indeed, the oxygen isotopic composition of E chondrite olivine, while somewhat variable, is consistent with derivation from the same isotopic reservoir as most other components of those meteorites (Weisberg et al. 2011). Olivine-bearing chondrules should open a window on the succession of events that led to their unique mineralogical assemblage.

Trace element geochemistry is a powerful tool to make further progress on the history of these objects. Indeed, how trace elements partition themselves among the different phases, in particular silicates, of chondrules is, given the wide variety of their geochemical behaviors, very sensitive to the formation conditions, and this at the order-of-magnitude level. Beside bulk analyses (e.g. Barrat et al. 2014; Lehner et al. 2014), previous trace element analyses on enstatite chondrite have included SIMS measurements of sulfides and in particular oldhamite (Crozaz and Lundberg 1995; Gannoun et al. 2011) as well as pyroxene (Weisberg et al. 1994; Hsu and Crozaz 1998). Given the low abundances of REE in olivine, Laser Ablation Inductively Coupled Plasma Mass Spectrometry (LA-ICP-MS), with its lower detection limits, is the method of choice for the olivine-bearing chondrules, and a technique which we have already successfully applied to chondrules in CR/CV carbonaceous chondrites (Jacquet et al. 2012) and LL ordinary chondrites (Jacquet et al. 2015). Here, we report *in situ* LA-ICP-MS analyses of the silicate phases in chondrules from enstatite chondrite Sahara 97096 (EH3), one of the most primitive unequilibrated enstatite chondrite known (Weisberg and Kimura 2012), focusing on olivine-bearing chondrules, in order to understand their similarities and differences with respect to their counterparts in other chondrite groups.

## *2. Samples and analytical procedures*

The polished section Sp1 of Sahara 97096 (EH3) from the meteorite collection of the Muséum National d'Histoire Naturelle de Paris (MNHN) was used in this study. Analytical procedures were similar to those of Jacquet et al. (2012) and are only briefly summarized here. The objects of interest were examined in optical and scanning electron microscopy (SEM—here a JEOL JSM-840A instrument). X-ray maps allowed apparent mineral modes of chondrules to be calculated using the

JMicrovision software (www.jmicrovision.com). Minor and major element concentrations of documented chondrules were obtained with a Cameca SX-100 electron microprobe (EMP) at the Centre de Microanalyse de Paris VI (CAMPARIS), using well-characterized mineral standards.

Trace element analyses of selected chondrules were performed by LA-ICP-MS at the University of Montpellier II. The laser ablation system was a GeoLas Q$^+$ platform with an Excimer CompEx 102 laser and was coupled to a ThermoFinnigan Element XR mass spectrometer. The ICP-MS was operated at 1350 W and tuned daily to produce maximum sensitivity for the medium and high masses, while keeping the oxide production rate low ($^{248}$ThO/$^{232}$Th ≤ 1%). Ablations were performed in pure He-atmosphere (0.65 ± 0.05 L•min$^{-1}$) mixed before entering the torch with a flow of Ar (≈ 1.00 ± 0.05 L•min$^{-1}$). Laser ablation condition were: fluences ca. 12J/cm² with pulse frequencies between 5 and 10 Hz were used and spot sizes of 26-102 μm, 51 μm being a typical value. With such energy fluences, depth speed for silicates is about 1 μm•s$^{-1}$. Each analysis consisted of 4 min on background analyses (laser off) and 40 s of ablation (laser on). Data reduction was carried out using the GLITTER software (Griffin et al. 2008). Internal standard was Si (Ca for augite), known from EMP analyses. The NIST 612 glass (Pearce et al. 1997) was used as an external standard. The absence of contamination by other phases was checked by examination of the time-resolved GLITTER signal, SEM imaging of ablation craters and comparison with EMP data. For each chondrule, a geometric (rather than arithmetic) averaging was used to calculate mean concentrations in olivine and pyroxene in order to minimize the impact of possible undetected contamination by incompatible element-rich phases. In the following text and in the figures, and unless otherwise noted, the data reported will be *chondrule means*, that is, for each phase, the average of the different (successful) analyses performed on that phase in a given chondrule.

## 3. Results

### 3.1 Petrography and mineralogy

In this section we provide a brief petrographic overview of analysed chondrules (whose properties are listed in Table 1), supplemented by selected images in Fig. 1. As in Jacquet et al. (2012), we call "chondrules" all silicate objects with petrographic evidence of melting regardless of their overall shape. We use standard chondrule classification: porphyritic, radial and barred types are denoted "P", "R" and "B" respectively, to which we add "O", "P" or "OP", depending on whether the olivine/(olivine+pyroxene) ratio is greater than 90 vol%, smaller than 10 vol% or intermediate. We call a "type I" chondrule a chondrule with Mg/(Fe+Mg) > 90 mol% in its ferromagnesian silicates, as are the vast majority of Sahara 97096 chondrules (with Fs$_{1-3}$, Fa$_{1-2}$).

Fourteen type I chondrules were analysed in Sahara 97096. RP, BP and PP (Fig. 1a-b) chondrules are the most abundant types (1 of each type was analysed by LA-ICP-MS), dominated by enstatite with interstitial mesostasis and silica pods. Although residual olivine chadacrysts may be seen in many chondrules, *bona fide* POP chondrules are rare and 6 were found to have crystals suitable for analysis (e.g. Fig 1c-f). These chondrules, where olivine crystals occur either independently in the mesostasis or poikilitically enclosed in enstatite, are richer in mesostasis (often devitrified) and accessory diopside than the former. They however differ significantly from their ordinary chondrite counterparts e.g. in the frequent occurrence of silica (e.g. near the margins or as pods in the mesostasis) or in the presence, if in smaller amount than in the bulk rock, of oldhamite and/or niningerite, along with troilite and kamacite (the latter two occurring as >10-100 µm inclusions or micron-sized intracrystal blebs).

Nearly monomineralic ferroan pyroxene (with $Fs_{8-17}$) -rich objects, often peppered with kamacite or troilite, were also analysed and include both porous spherules (Fig. 1h) and irregularly shaped clasts with vermicular silica, often surrounded by radial cracks (Fig. 1g). 2 and 3 objects were analysed for each type, respectively.

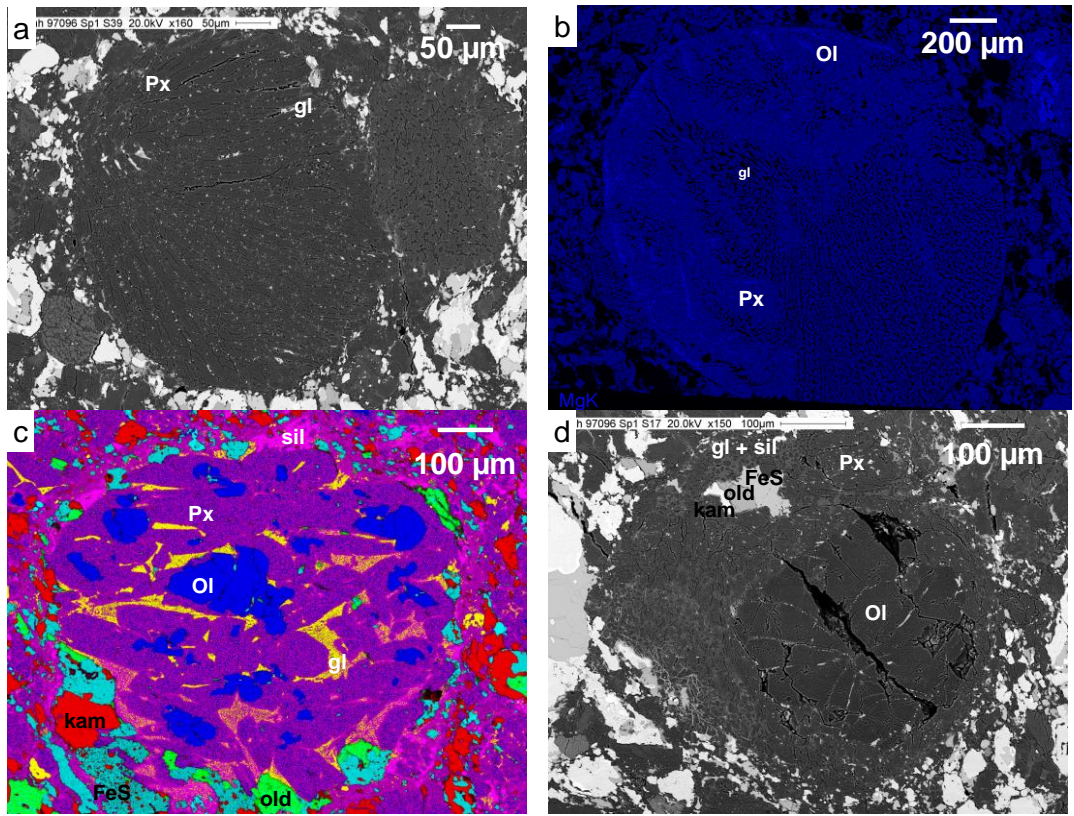

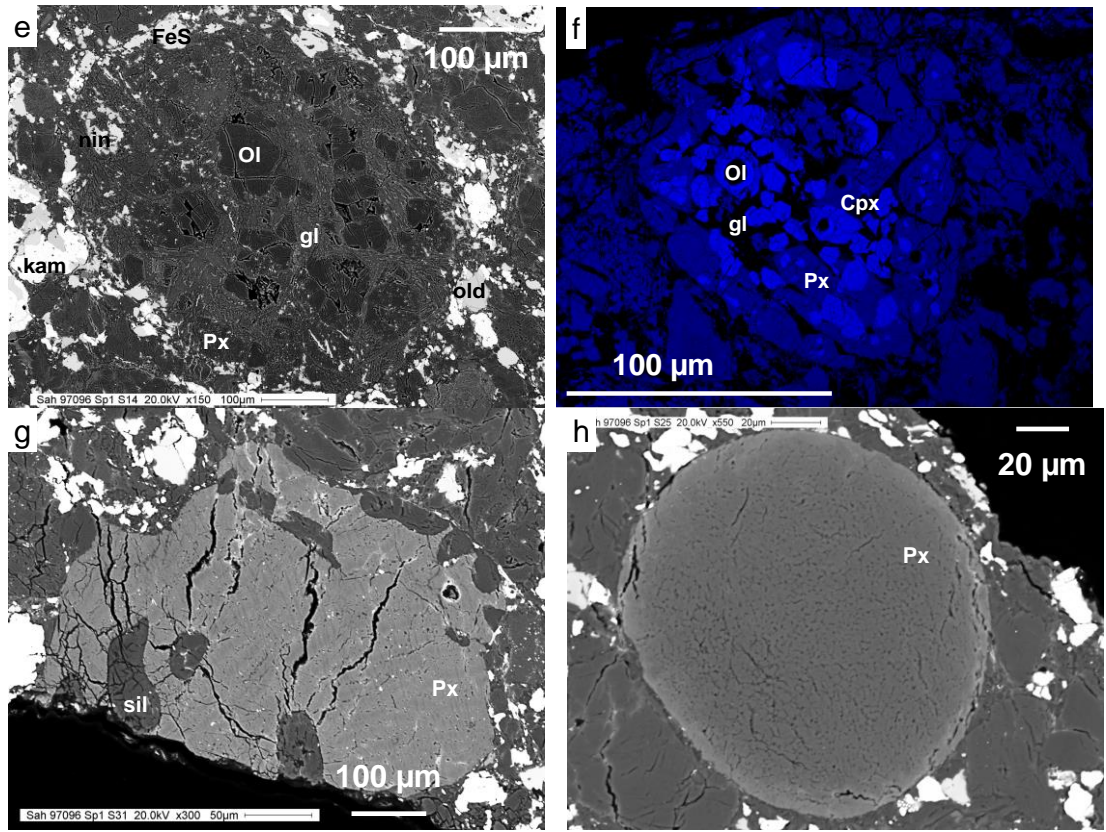

**Figure 1**: Images (back-scattered electrons unless otherwise noted) of selected chondrules and other inclusions of Sahara 97096 analyzed in this study. (a) Radial pyroxene chondrule S39, peppered with sulfide. (b) Mg X-ray map of porphyritic pyroxene chondrule S2, with euhedral enstatite crystals (sometimes adopting a barred texture) and interstitial mesostasis. Patches of oldhamite occur as well as rare <30 µm thick olivine streaks near the border. (c) Combined X-ray map of porphyritic olivine pyroxene chondrule S9. Rounded olivine grains are partially or entirely enclosed in enstatite laths, with interstitial finely devitrified mesostasis. An opaque shell of kamacite, oldhamite, troilite surrounds the lower left side of the chondrule. Blue = Mg, pink = Si, red = Fe, green = Ca, yellow = Al, cyan = S. (d) POP chondrule S17 with a large 300 µm olivine crystal exhibiting curvilinear troilite trails. The crystal is overgrown by enstatite grading in a finer-grained texture with interstitial mesostasis and silica pods. A Si-bearing kamacite/oldhamite/troilite association prominently occurs on the upper left. (e) POP chondrule S14 with euhedral grains of olivine partly resorbed in enstatite, embedded in a glassy mesostasis with abundant augite crystallites. (f) Mg X-ray map of porphyritic olivine pyroxene chondrule S16, with rounded olivine crystals and large enstatite laths overgrown by augite, and glassy mesostasis. Round holes are LA-ICP-MS ablation craters. (g) Inclusion S31 of bronzite (with fine iron-poorer lamellae) poikilitically enclosing vermicular silica inclusions surrounded by radial cracks. (h) Porous pyroxene spherule S25, with a slight Fe enrichment in the outermost microns. Abbreviations: Ol = olivine, Px = low-Ca pyroxene, Cpx = Ca-rich pyroxene, gl = mesostasis, kam = kamacite, FeS = troilite, old = oldhamite, sil = silica.

## 3.2 Chemistry

Trace element data for each chondrule type and phase are summarized in Table 2, and individual chondrule data are provided in the Electronic Annex. Rare earth element (REE) and other lithophile element patterns are shown in Fig. 2-4 (with comparison to LL chondrites; cf Jacquet et al. 2015). We now describe the chemical composition of the chondrules, starting with type I chondrules (excepting two with anomalous results, discussed afterward), with particular emphasis on REE. We denote by LREE, MREE and HREE the light, middle and heavy REE, respectively. Normalizations to CI chondritic abundances, when mentioned, will be made using the Lodders (2003) values and denoted by a "N" subscript.

The REE pattern of mesostasis exhibits a flat baseline (e.g. $2 \leq Lu_N \leq 20$; Fig. 2a) with common negative anomalies in Eu (Eu/Eu* = 0.02-0.5; with $Eu^*_N = (Sm_N+Gd_N)/2$) and Yb (Yb/Yb* = 0.08-1.0 with $Yb^*_N = (Er_N+Lu_N)/2$), and one Sm anomaly in S19 ($(Sm/Nd)_N = 0.1$). While we are unaware of trace element analyses of enstatite chondrite chondrule mesostases and Yb anomalies have not been encountered in our previous studies on ordinary and carbonaceous chondrites, such features are consistent with bulk chondrule analyses (see below) as mesostasis carries most of the incompatible element budget of chondrules. Mesostasis is generally comparable to its LL type I chondrule counterparts, although somewhat depleted in many minor elements (Fig. 4), with large depletions in Ca (1.5 vs. 7.3 wt%) and in P (8 vs. 800 ppm) but not in Na (6 vs. 3 wt%).

Enstatite ($Fs_{1-3}$) displays REE patterns with steady enrichment from the LREE ($0.03 \leq La_N \leq 0.06$; Fig. 2a) to the HREE ($0.2 \leq Lu_N \leq 0.4$) with sometimes weak negative Eu anomalies (Eu/Eu* = 0.3-0.8). Apart from this one, the patterns are intermediate between the "type II" and "type III" patterns of Hsu and Crozaz (1998), who analysed enstatite in Qingzhen and Yamato 691 (EH3), although *bona fide* examples of their "type III" patterns (with very depleted LREE around $10^{-3}$ x CI) have not been encountered, but this may be due to low statistics in this olivine-centered study. (On the other hand, we certainly encountered their "type I" patterns (flat), which were however systematically found to be due to contamination with mesostasis, as Hsu and Crozaz (1998) had already interpreted for theirs, and therefore rejected).

As with other chondrite groups, olivine ($Fa_{1-2}$) has the lowest REE abundances of the analysed phases, with patterns paralleling that of enstatite ($0.006 \leq La_N \leq 0.01$; $0.07 \leq Lu_N \leq 0.3$; Fig. 2a). Similar to our study on LL chondrites, no very LREE-depleted olivine, such as the few ones described by Jacquet et al. (2012) in Vigarano and Renazzo, has been encountered. Negative Eu anomalies occur (Eu/Eu* = 0.1-1).

Two type I chondrules, S14 and S16 yielded anomalous results (Fig. 3): Olivine in S14 (Fig. 3a) and S16 (Fig. 3b) has a LREE-enriched pattern ($La_N$ = 0.4 and 0.3; $Lu_N$ = 0.2 and 0.1), possibly reaching a minimum around Er, with S16 olivine exhibiting positive Sm and Yb anomalies (Sm/Sm* = 2; Yb/Yb*= 4). Same holds for enstatite in chondrule S16 with $La_N$ = 1 and $Lu_N$ < 0.1, with positive Sm and Yb anomalies interrupting the concave baseline (Sm/Sm*= 3 with $Sm^*_N = (Nd_N+Eu_N)/2$ and Yb/Yb* > 20), although augite there has a "typical" augite REE pattern, rising from La = 4 x CI to Sm = 10 x CI and flattening for the HREE, with a pronounced negative Eu anomaly (Eu/Eu*=0.06). Mesostases in both chondrules both have negative Eu and Yb anomalies but that in S16 has a deep negative Sm anomaly ($((Sm/Nd)_N < 0.02)$.

Bulk analyses were also performed on a RP chondrule S39 and a radial/barred textured portion of PP chondrule S2, yielding roughly flat REE patterns (Fig. 2b) around chondritic levels with Eu and Yb anomalies (positive for S2, negative for S39, which also has a negative Sm anomaly ($(Sm/Nd)_N$=0.4)).

This is consistent with the average chondrule composition for Sahara 97072 (paired with Sahara 97096) given by Lehner et al. (2014).

Ferroan pyroxene ($Fs_{10-17}$) from three pyroxene-silica clasts (S31, S33, S38) was analysed, yielding flat REE patterns ($La_N$ = 0.7-2; Fig. 2c), with S31 and S38 exhibiting negative Eu anomalies (Eu/Eu* = 0.1). Ferroan pyroxene ($Fs_{8-10}$) from porous spherules S25 and S40 exhibit LREE/HREE enrichment ($La_N$ = 0.5 and 0.3; $Lu_N$ = 0.2 and 0.04; Fig. 2c) with positive Ce anomalies (Ce/Ce* = 2 and 3 with $Ce^*_N$ = ($La_N+Pr_N$)/2). The former are comparable to the fairly flat black pyroxene patterns of Weisberg et al. (1994), although these tended to be somewhat LREE-depleted.

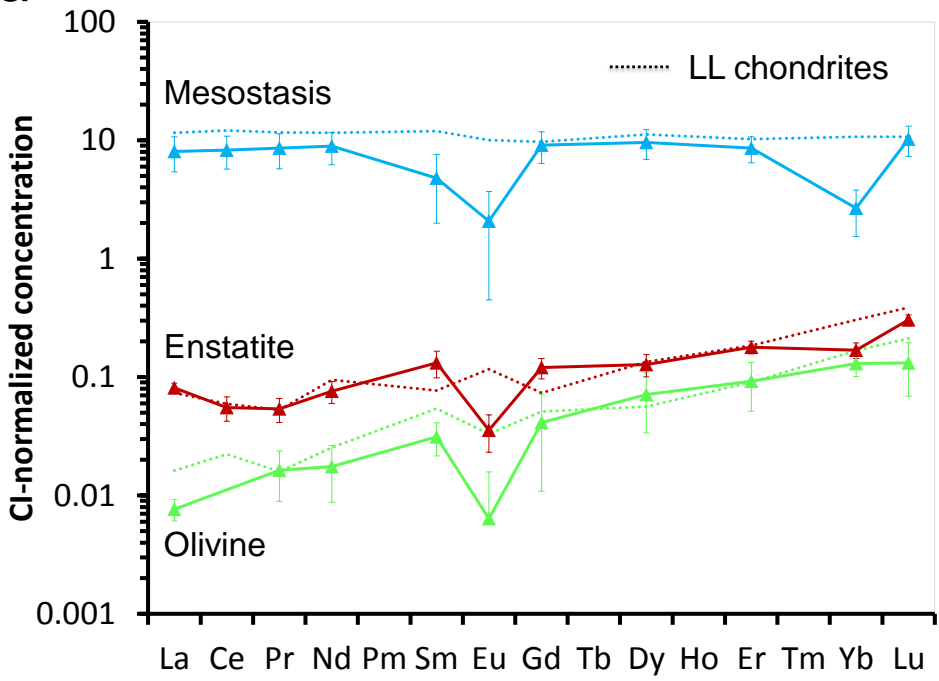

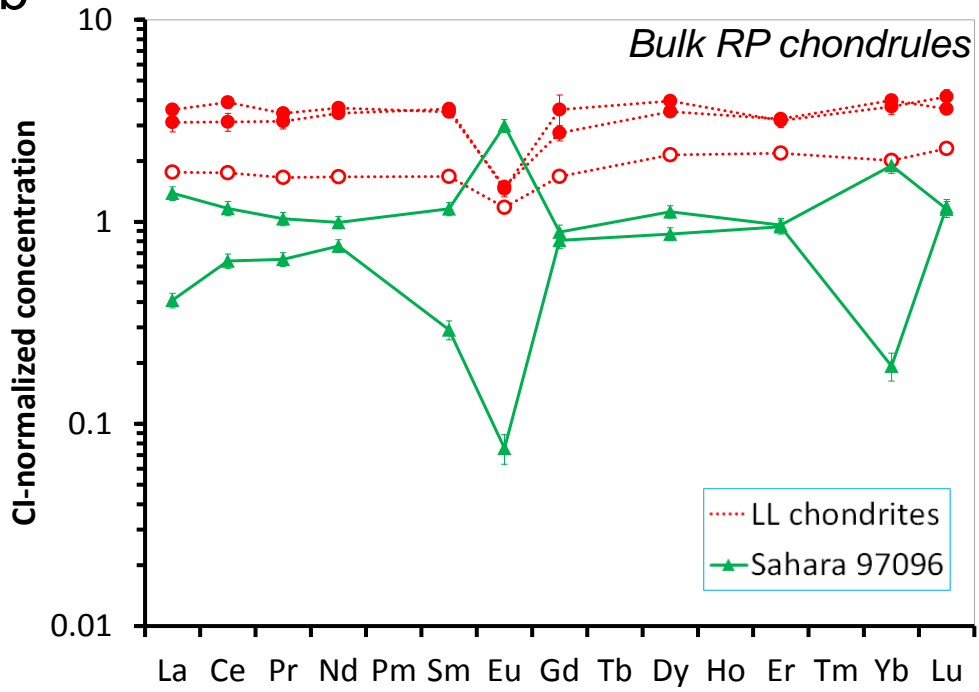

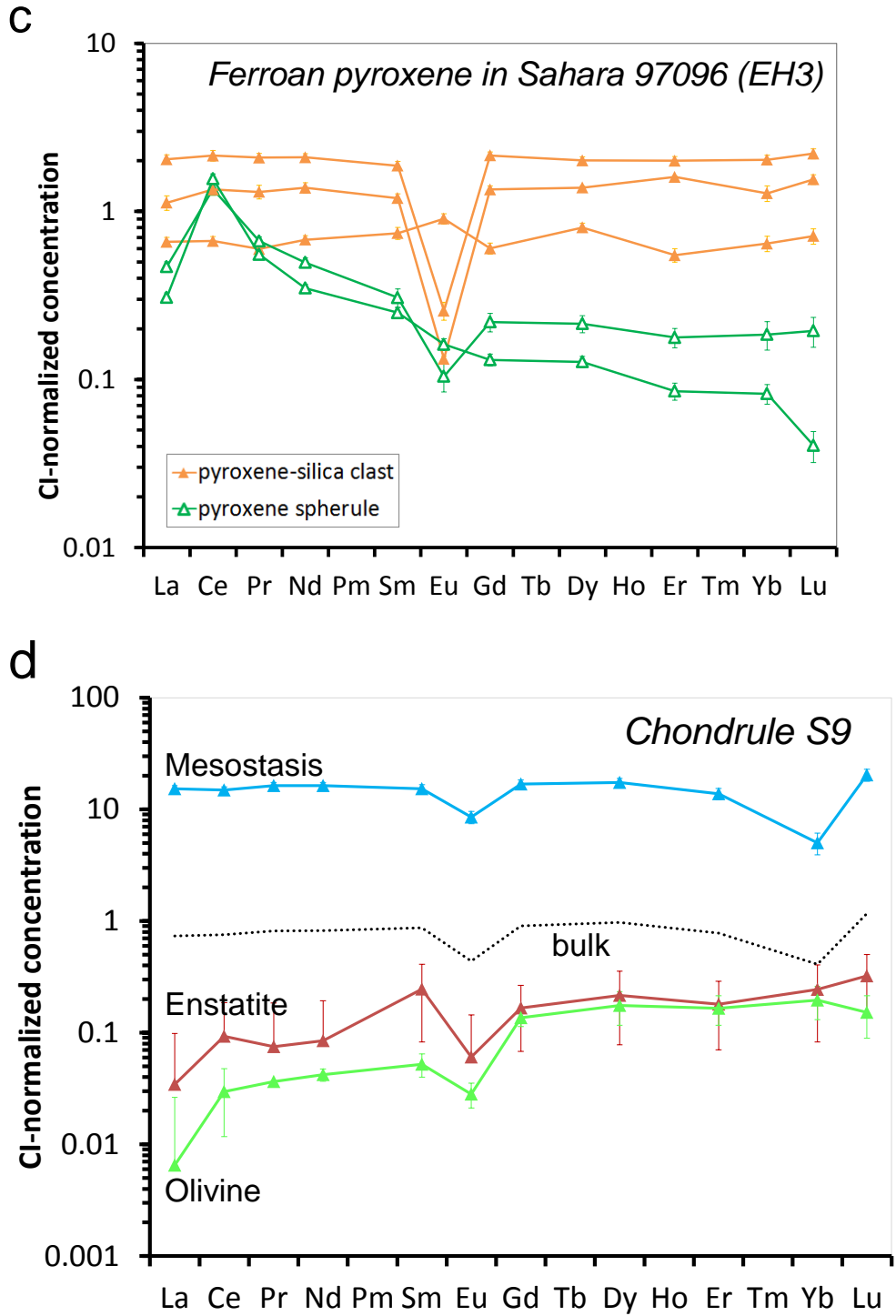

**Figure 2**: (a) CI-normalized rare earth element patterns of Sahara 97096 chondrule silicates and mesostasis (averages of 10 (olivine), 13 (enstatite) and 7 (mesostasis) analyses; excluding anomalous LREE-enriched ferromagnesian silicates, see Fig. 3). Data for LL chondrite type I chondrules are also shown in dotted lines for comparison. (b) Bulk REE patterns of radial pyroxene chondrules. (c) Pyroxene REE pattern in ferroan pyroxene-bearing objects in Sahara 97096, divided in pyroxene spherules (green) and pyroxene-silica clasts. (d) REE patterns of silicate phases in a single chondrule (S9) and reconstructed bulk. Error bars represent one standard error of the mean.

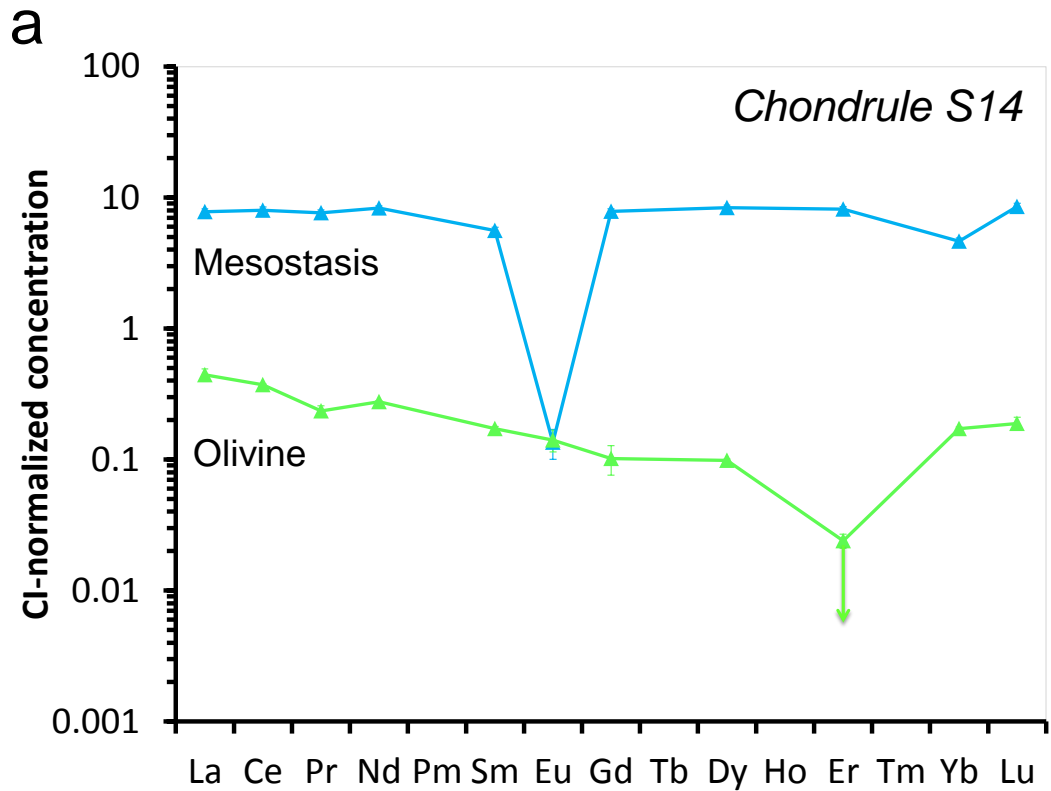
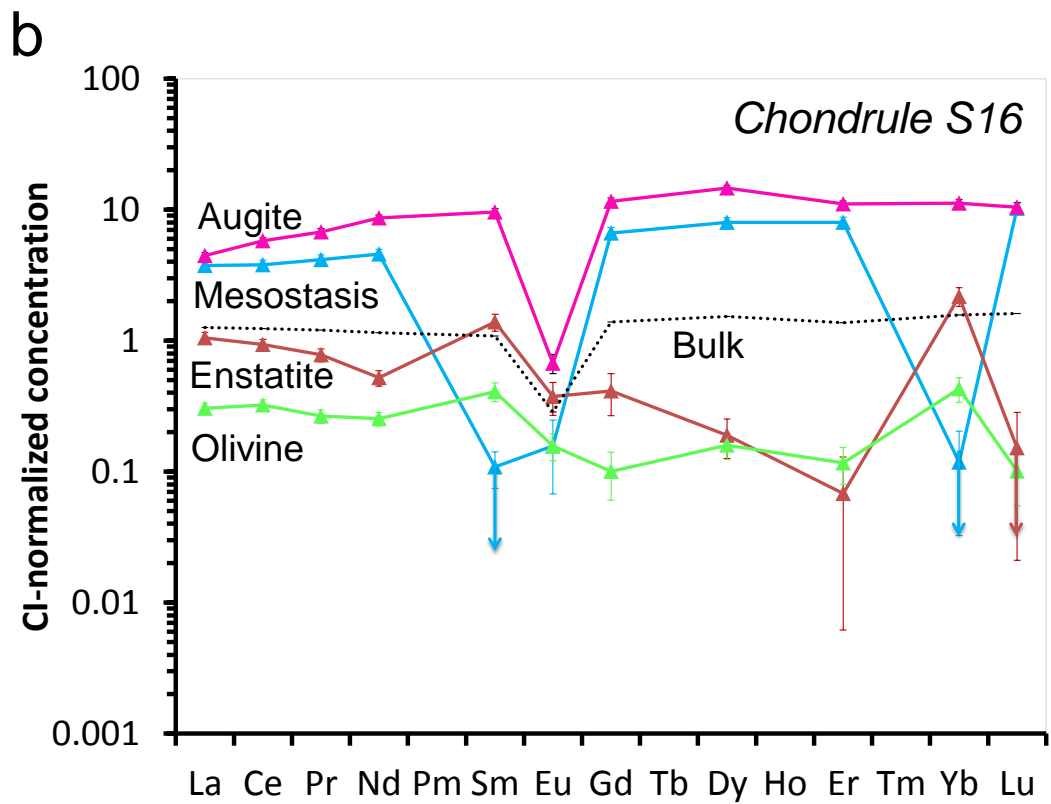

**Figure 3**: REE patterns of phases in chondrules with anomalous LREE-enriched ferromagnesian silicates, viz. chondrule S14 (a) and chondrule S16 (b). Downward arrows indicate upper limits and error bars are one standard deviation.

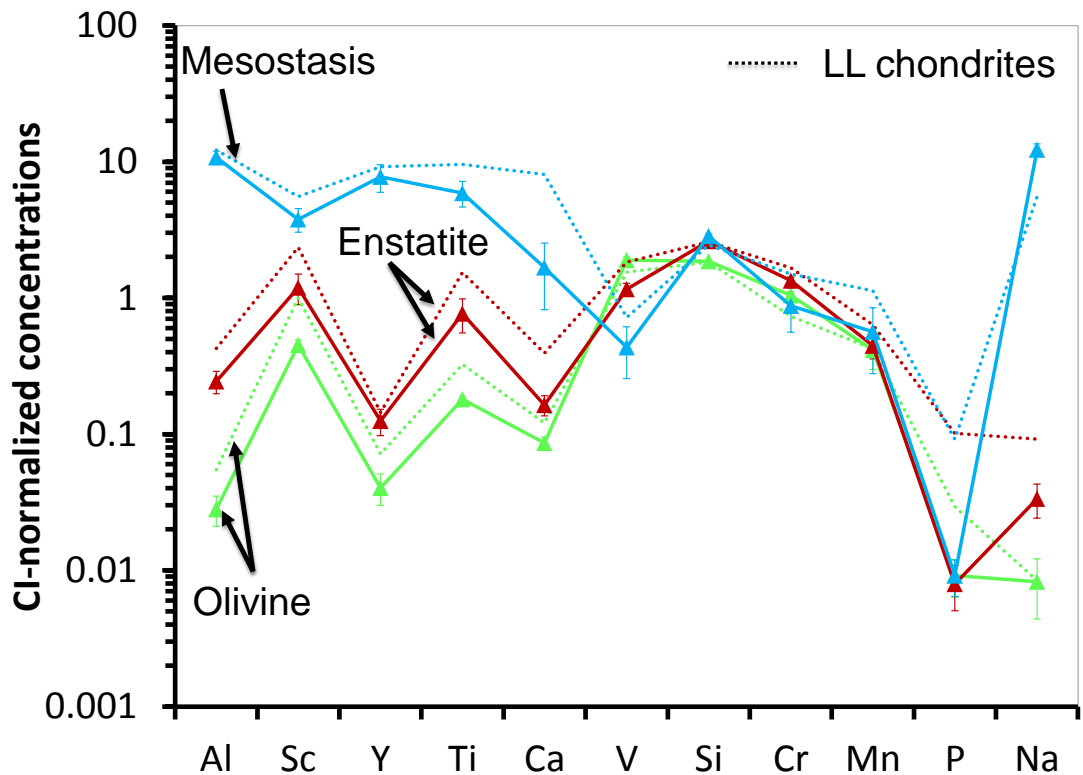

**Figure 4**: CI-normalized average concentration of selected elements arranged in order of increasing volatility for normal olivine, low-Ca pyroxene and mesostasis. Patterns for type I chondrules in LL chondrites are shown in dotted lines for comparison. Error bars represent one standard error of the mean.

## *4. Discussion*

### 4.1 Origin of normal olivine-bearing chondrules

  In this section, we set aside the chondrules S14 and S16 whose abnormal negatively sloped REE patterns will be discussed in the next section. As we have seen, silicates in the other "normal" chondrules have lithophile element patterns comparable to those of type I chondrules in other chondrite groups, and consistent with an igneous origin (Alexander 1994; Jones and Layne 1997; Ruzicka et al. 2008; Jacquet et al. 2012, 2015). The patterns of REE partition coefficients indicated by both olivine and pyroxene may be understood in terms of a conventional lattice strain model (Wood and Blundy 2003; Nagasawa 1966), where the crystallographic site to be substituted by the REE has a radius of about 0.7 Å in both cases (Wood and Blundy 2003; Bédard 2005; Bédard 2007), whereas REE have (trivalent) ionic radii ranging from 1.03 Å for $La^{3+}$ to 0.86 Å for $Lu^{3+}$ (sixfold coordination, high-spin; Li 2000). Thus LREE, whose sizes are furthest from that of the substitution site, partition less easily than HREE in either ferromagnesian silicate, as can be seen in the Onuma diagrams (Onuma et al. 1968) plotting partition coefficients as a function of ionic radius in Fig. 5 (apparent overabundances of LREE relative to a smooth lattice strain expectation will be also discussed in the next section). Cartier et al. (2014) found that the enstatite/melt partition coefficients for REE did not depend on oxygen fugacity for a wide range of reducing conditions (from IW-8.2 to IW-0.8 where "IW" stands for the Iron-Wüstite buffer).

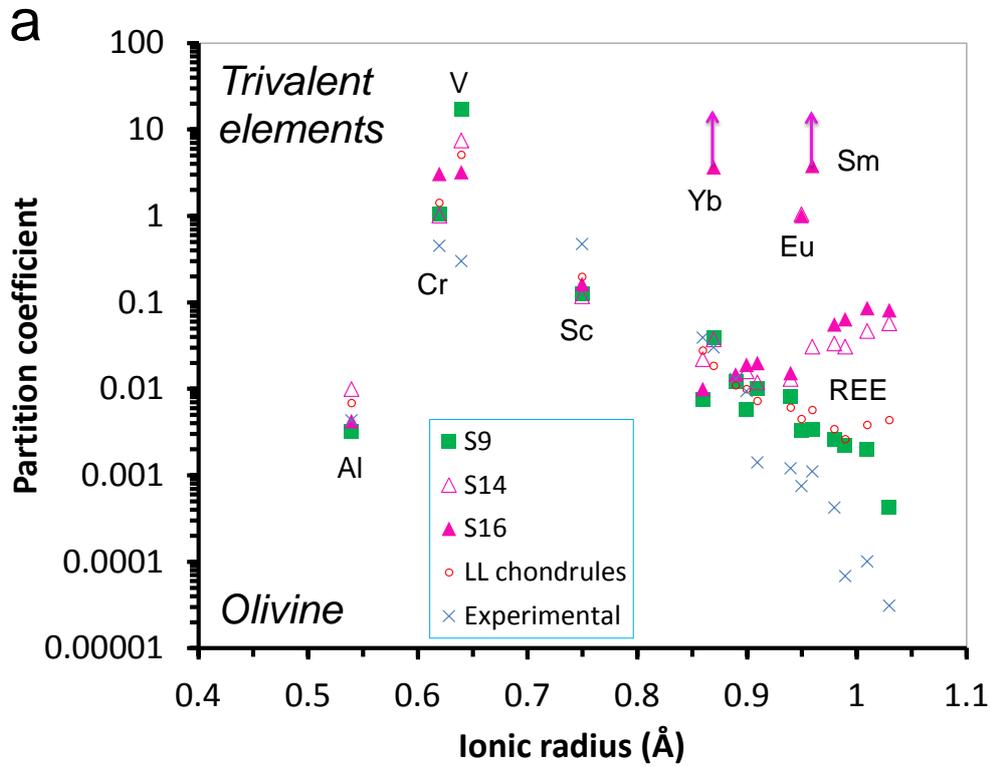

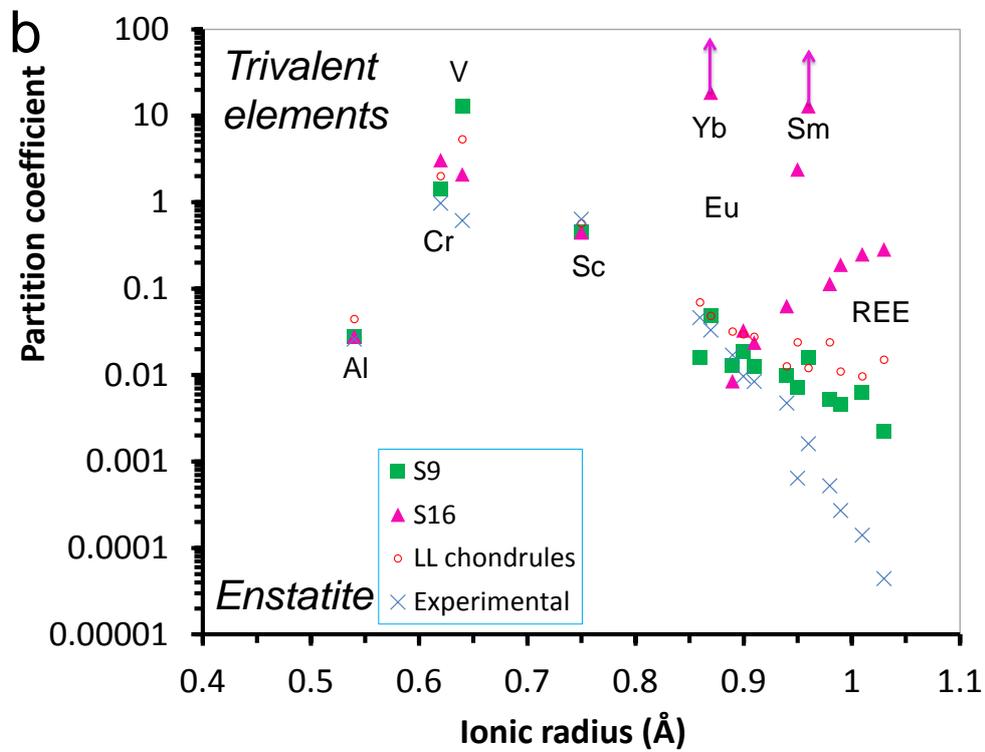

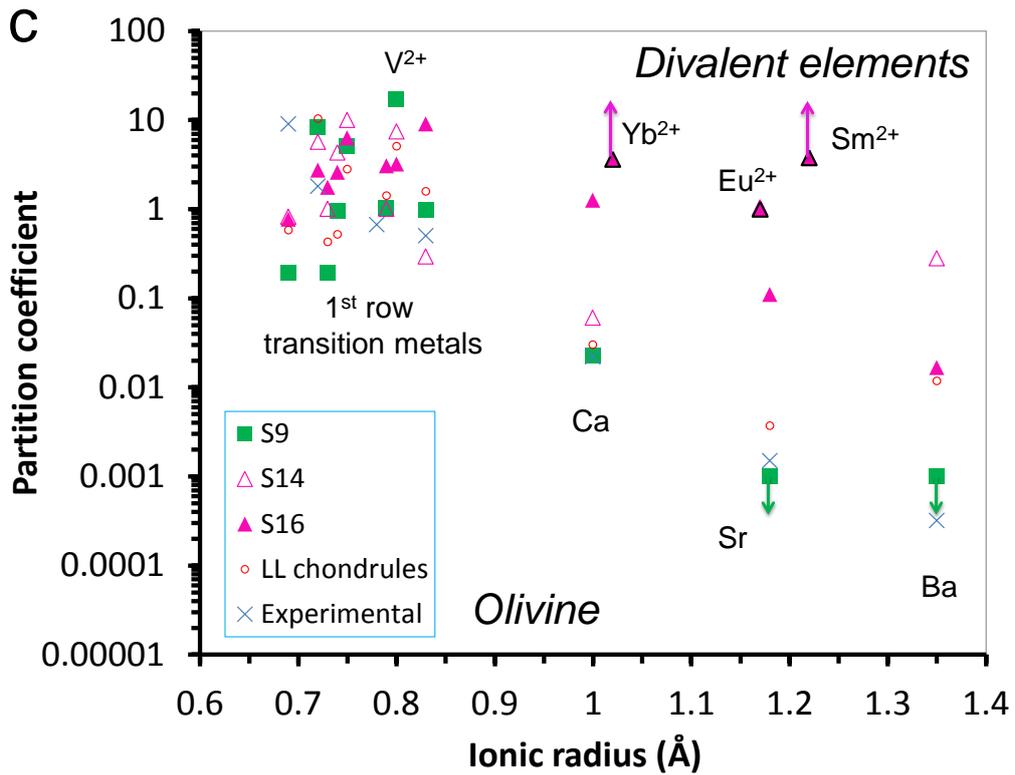

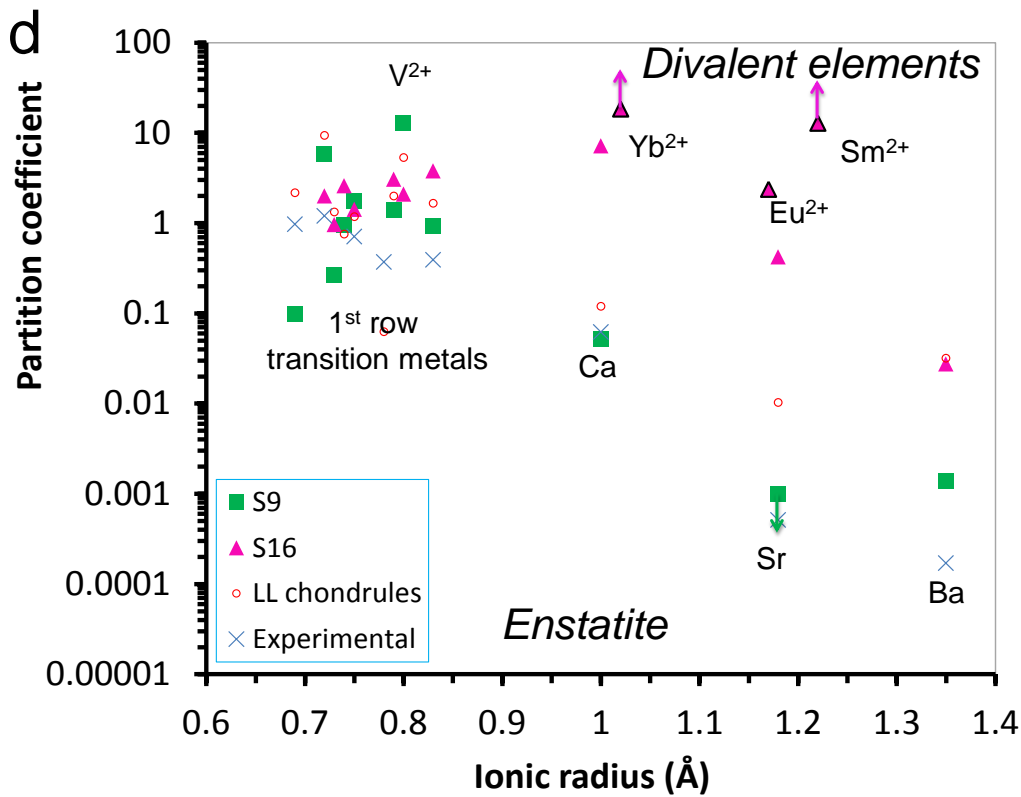

**Figure 5**: Onuma diagrams (i.e. silicate/mesostasis partition coefficient versus ionic radius) for trivalent (a,b) and divalent (c,d) elements and for olivine (a, c) and enstatite (b, d). Plotted are data for a normal POP chondrule (S9), the two anomalous chondrules S14 and S16, as well as the average of LL chondrite type I chondrule (Jacquet et al. 2015) and experimental values (Kennedy et al. 1993; runs PO49 for olivine and RPII45 for enstatite) for comparison. Ionic radii (sixfold coordination, high-

spin when relevant) are taken from Li (2000). A maximum near 0.7 Å is apparent for all silicates, in agreement with the lattice strain models, but anomalous LREE enrichments mimic the effect of another substitution site (near 1 Å according to the panels for trivalent elements). The data of putatively divalent REE (Eu, Yb, Sm) in chondrule S16 are also plotted and specially identified on the Onuma diagrams for divalent ions. Arrows signal lower or upper limits.

The analyzed olivine and enstatite do not seem to record much more reducing conditions than their counterparts in type I chondrules in ordinary chondrites. Fe contents are 0.8-1.2 and 0.3-1.4 wt% in olivine and enstatite, respectively (see also Weisberg and Kimura 2012), compared to 0.5-5 wt% and 0.5-2.7 wt% in the type I chondrules we analysed in LL3 chondrites (Jacquet et al. 2015). Interestingly, Simon et al. (2013) reported the presence of $Ti^{4+}$ in olivine and enstatite in enstatite chondrites, also a sign of oxidizing conditions. Also, Cartier et al. (2014) showed that the ratio D(Cr)/D(V) (where D denotes the enstatite/melt partition coefficients) could be used as a proxy for oxygen fugacity. Assuming the melt Cr/V ratio to be equal to the mean bulk chondrule compositions (prior to further element redistribution), which is 0.8 x CI for either olivine-bearing chondrules in the Qingzhen EH3 chondrite (Grossman et al. 1985) or porphyritic chondrules in unequilibrated ordinary chondrites (Grossman and Wasson 1983), we obtain (rough) ranges of 0.9-2.0 and 0.7-1.3 for our Sahara 97096 and Bishunpur/Semarkona analyses, respectively, in either case indicating oxygen fugacities between IW-2 and IW+1 (Cartier et al. 2014).

In Figure 6a, we calculate the melt composition in equilibrium with olivine in the different chondrule analyzed. As in our earlier work on type I chondrules in ordinary and carbonaceous chondrites (Jacquet et al. 2012, 2015), these compositions are too enriched in refractory lithophile elements to match bulk chondrules (see Lehner et al. 2014; Fig. 2b), as in a fractional crystallization scenario, and appear more consistent with a mesostasis composition. This, as previously, indicates a batch crystallization scenario where olivine maintained equilibrium with the melt during precipitation of the silicates, which may indicate cooling timescales >10 h (Jacquet et al. 2012, 2015). This is tested in Fig. 6b where the olivine/mesostasis partition coefficients are compared to experimental values by Kennedy et al. (1993). However, while some elements indeed hover near the predicted equilibrium values, several anomalies are visible. Some may simply reflect the change of valence state of some elements under conditions more reducing than those of the experiments chosen (e.g. Ti, V, Cr), affecting the partition coefficients, as mentioned previously (see also Ruzicka et al. 2008; Jacquet et al. 2012). But other anomalies e.g. in Ca, Mn, P, Na—which show very different concentrations in Sahara 97096 chondrule mesostases than in LL chondrite chondrules—suggest that the composition of the mesostasis has evolved out of equilibrium with the olivine.

This is especially borne out by rare earth elements. Indeed, the mesostasis shows frequent deep negative anomalies Eu and Yb uncorrelated with those in the coexisting silicates (Fig. 7). This contrasting situation is unlikely to arise from changed partition coefficients due to Eu and Yb being possibly divalent; not only would some fine-tuning between such partition coefficients and bulk composition be required to coincidentally cancel any significant anomalies in olivine and enstatite, but as to Eu, this would go in the wrong direction, as $Eu^{2+}$ would be *more* incompatible than $Eu^{3+}$ in either silicate (e.g. Cartier et al. 2014). So clearly, mesostasis and silicates cannot be (presently) in equilibrium and the latter have to be relict in some sense, as also suggested by frequent "dusty" reduction textures (e.g. Rambaldi et al. 1983).

For objects like S17 (Fig 1d) where a large single olivine phenocryst is set in a finer-grained groundmass (see also object C11 in Weisberg et al. (2011) whose olivine was found to have R chondrite-like oxygen isotope signature), olivine may conceivably be xenocrystic (that is, be genetically unrelated to the host), and in such case the anomalies of the "host" material could conceivably be inherited from the precursor. Specifically, the host material precursor could have comprised group III CAI (Mason & Taylor 1982), or a condensate formed at supersolar C/O ratios as proposed by Pack et al. (2004) in the case of a Sm anomaly (like chondrule S19 mesostasis, or bulk chondrule S39; see also Lehner et al. (2014) and the anomalous ordinary chondrite chondrule analyses of Pack et al. (2004)). A precursor effect was also proposed by Hsu and Crozaz (1998) to explain Eu and Yb anomalies in part of their EL3 chondrule enstatite analyses, although the accompanying mesostasis was not analysed, so it is not possible to unambiguously distinguish from fractionations during chondrule formation for their analyses. But given that whenever chondrules contain several silicate grains, all these give similar REE patterns, it seems difficult to envision that they are all xenocrysts, and it thus seems that the mesostasis composition *evolved* from relatively flat REE patterns to the anomalous present-day ones.

How could such an evolution of the melt composition come about? Schematically, the melt could either have *gained* a component with negative Eu, Yb and/or Sm anomalies, or *lost* one with positive anomalies in those elements. The "gain" option is quite unlikely given the rarity of CAIs (a fortiori with group III signatures) or any non-chondrule object with the wished anomalies (this in fact is also a general problem for a (non-chondrule) precursor effect picture). As to the "loss" option, one could first envision evaporation of Eu, Yb and Sm which are more volatile (in terms of absolute condensation temperature but also compared to the other REEs) under reducing conditions (Boynton 1989; Lodders and Fegley 1993). It is however difficult to envision evaporation of REE with e.g. Na and Si still being retained in large proportions in the chondrules (Fig. 2d; Grossman et al. 1985). But a more promising component to consider may be oldhamite.

Indeed, oldhamite is not only the major REE carrier in enstatite chondrites, it also frequently presents positive Eu and/or Yb anomalies, especially in Sahara 97096 (Gannoun et al. 2011) where 97 % of the grains belong to the type C-D of Crozaz and Lundberg (1995), with relatively flat REE baselines at 10-100 x CI, enrichment in LREE, and, indeed, positive Eu and Yb anomalies. The conventional interpretation of this REE enrichment is that oldhamite is an early condensate formed under reducing conditions — in which case it would offer little prospect to relate to the evolution of chondrule composition —; yet even notwithstanding the absence of expected collateral condensates such as silicon carbide (see e.g. Lehner et al 2013), the positive anomalies in Eu and Yb, which should be *volatile*, are problematic (Hsu and Crozaz 1998; Gannoun et al. 2011). To be sure, one could still construct a scenario where the early, REE-enriched but Eu, Yb, Sm-depleted condensates were removed followed by further condensation of oldhamite inheriting the Eu, Yb, Sm positive anomalies of the residual gas (Lodders and Fegley 1993), similar to the interpretation of group VI CAIs. However, bulk enstatite chondrites, and Sahara 97096 in particular, show fairly unfractionated REE patterns (Barrat et al. 2014), and, again, show little in situ evidence for the purported complementary (or transitional, see Lodders and Fegley 1993) condensates, especially in the oldhamite population itself.

Here we entertain the alternative hypothesis that at least part of the oldhamite in enstatite chondrites has an igneous origin and crystallized during chondrule formation. In fact, while most oldhamite grains occur outside chondrules, analyses of 5 intrachondrule oldhamite grains by Crozaz and Lundberg (1995) also assigned 3 of them to type C-D (with the last 2 being of type A, i.e. with REE uniformly enriched at ~100 x CI). This is noteworthy as, despite its high melting point as a pure phase, oldhamite should not survive melting temperatures above 1000°C in such assemblages (McCoy et al. 1999). So unless the analysed oldhamite grains were ingested late in the cooling phase of the chondrule and escaped resorption, the type C-D signature must be igneous in these cases. At first sight, a problem with such a possibility would be that experimentally measured oldhamite/silicate melt partition coefficients rarely exceed 5, and never 20, for REE (Dickinson & McCoy 1997). In fact, a comparable paradox is presented by similarly REE-rich oldhamite in aubrites (e.g. Wheelock et al. 1994), whose petrogenesis would leave even less chance for the survival of any nebular relict. Dickinson and McCoy (1997) however proposed that part of the REE enrichment of aubritic oldhamite could be due to late crystallization in equilibrium with an evolved (incompatible-element enriched) liquid, and this is exactly in the sequence of event we are led to envision for our chondrules. Equilibration of oldhamite with chondrule *mesostasis*, enriched by one order of magnitude in REE above chondritic values, would indeed offer some leverage to account for the absolute magnitude of REE concentrations in EC oldhamite. One would however have yet to explain the positive Eu and Yb anomalies. To that end, we note that while we are not aware of any attempt to rationalize oldhamite/melt partition coefficients in terms of a lattice strain model — Lodders (1996) offer a thermodynamic treatment assuming ideal solution but finds limitations with this assumption —, such a parameterization seems likely applicable as oldhamite has a NaCl structure. Indeed KCl was one of the first two phases originally investigated experimentally and modelled in a lattice-strain-like framework by Nagasawa (1966). Then the positive Eu anomaly could be due to Eu being in the divalent form (as seen in the experimental partitioning data; Lodders 1996; Dickinson and McCoy 1997), possibly exactly fitting the substitution site size similarly to the case of calcic plagioclase; hence perhaps the relative independence of Eu concentration to the other REE (Gannoun et al. 2011), as the partitioning of the latter may have been possibly more dependent on the varying stiffness of the site (e.g. as a function of temperature). The Yb anomaly would likewise be attributable to Yb being in divalent form — implying reduced conditions at that stage —, and its correlation with La noted by Gannoun et al. (2011) may be linked to the proximity of the radius of $Yb^{2+}$ to that of $La^{3+}$ (1.02 vs. 1.03 Å; Li 2000), although it should be cautioned that ion charge also matters in lattice strain parameterization (Wood and Blundy 2003). Obviously, more experimental oldhamite/melt partitioning work on a set of elements more extended than REEs, as well as more analytical data on chondrule-hosted oldhamite would be necessary to test these ideas.

This being said, how would oldhamite fit in the thermal history of our chondrules? We follow Lehner et al. (2013) and experimental work by Fleet and MacRae (1987) in advocating a moderate-temperature (1400-1600 K; Lehner et al. 2013) sulfidation event, in a S-rich gaseous reservoir, at the origin of niningerite and oldhamite in enstatite chondrite chondrules. This is also supported by high S concentrations in chondrule mesostases measured by Piani et al. (2013). We note that less extensive sulfidation events have also been proposed for type I chondrules in CV chondrites (Marrocchi and Libourel 2013) and the ungrouped chondrite Kakangari (Berlin et al. 2007) — whose chondrules, incidentally, have oxygen isotopic compositions close to those of enstatite chondrites (Nagashima et al. 2011). Except for clear-cut xenocrystic grains, we cannot decide for certain whether this sulfidation event is independent of the crystallization of the pre-existing ferromagnesian silicates (e.g. Rambaldi et al. 1983) or actually forms with it a single sequence of events (similar to the two-stage scenario (olivine/pyroxene) preferred by Jacquet et al. (2013) for CR chondrite chondrules), with a change in redox conditions. Oldhamite crystallized during this event would have fractionated REE by igneous partitioning with the mesostasis as mentioned above, and would have largely been expelled from the chondrules, which are indeed depleted in opaque minerals and the accompanying siderophile/chalcophile elements (Grossman et al. 1985). Loss of those opaques, which is quite general in chondrules across the different chondrite groups (Grossman and Wasson 1985; Campbell et al. 2005; Jacquet et al. 2013), could have come about through surface tension effects (King 1983; McCoy et al. 1999; Uesugi et al. 2008) or inertial accelerations given density differences with the silicate melts (e.g. with troilite or metal), as witnessed for cosmic spherules (Genge and Grady 1998). Whatever that may be, Grossman et al. (1985) found evidence for oldhamite fractionation in Qingzhen (EH3) bulk chondrule compositions in the form of intercorrelations between Ca, Eu and Se, to which perhaps the overall Ca depletion of the analysed mesostases may be also be traced back. It may be commented that Grossman et al. (1985) did not then favour fractionation *during* chondrule formation on the ground that S should rapidly volatilize (and Defouilloy et al. (2013) saw little isotopic fractionation of S in enstatite chondrites). Nevertheless, recent evidence of such retention of volatile elements like Na despite prolonged thermal processing (Alexander et al. 2008; Hewins et al. 2012) indicates that, however mysterious the actual reason, this must not have specifically been an issue. The abundant opaque nodules in the matrix of EH chondrites may then have been at least partly derived from chondrules, and some complementarity is indeed indicated by the smooth REE patterns of the bulk meteorites (e.g. Ebel et al. 2015), despite negative and positive anomalies in Eu, Yb and/or Sm in the chondrule and oldhamite populations, respectively, as well as fairly chondritic overall bulk abundances in many siderophile and chalcophile elements (e.g. Barrat et al. 2014; Lehner et al. 2014). Although our geochemical focus on REE have led us hitherto to consider solely oldhamite, the opaque nodules of course generally contain other sulphides, metal, schreibersite etc. often in complex intergrowth (e.g. Gannoun et al 2011). Indeed, McCoy et al. (1999) report that partial melting experiments on Indarch (EH4) give rise to 3 immiscible metallic melts (Si, P and C-rich respectively) and 2 immiscible melts, a Fe- and, in particular, a FeMgMnCa-rich ones, so it is certainly conceivable that opaques were expelled as mixtures (sometimes still present in chondrules, e.g. Fig. 1d), whether still homogeneous or already in the process of phase separation, as in fact specifically suggested by McCoy et al. (1999) themselves. This may perhaps account for the depletion in Mn and P (now carried by niningerite and schreibersite, respectively) of the mesostasis, in addition to that in Ca (Fig. 4). Our hypothesis does not prohibit further evolution of the nodules after ejection; in particular they may have further interacted with the gas (e.g. $H_2S$), or collided with each other (see e.g. Fig. 3e,f of Gannoun et al. (2011) for a frozen-in oldhamite/kamacite compound and an isolated oldhamite fragment, respectively), as may be perhaps necessary to account for layered textures or even the presence of silicates whose chemistry Ikeda (1989b) found consistent with a chondrule origin.

Finally, we note that Lehner et al. (2013) proposed that since Mg would be made more volatile in the purported sulfidation events, its loss could explain the low Mg/Si ratios of enstatite chondrites — assuming that Mg never recondensed. We point out that Mg/Si fractionation is a general phenomenon among non-carbonaceous chondrites, with a distinct trend radiating from CIs in the Mg/Al vs Si/Al diagram (see e.g. Larimer & Wasson 1988), so may require a more general mechanism (see e.g. Jacquet (2014) for more discussion), even though Mg evaporation may have contributed to it. Also, no Mg isotopic difference between different chondrite groups has been resolved (Teng et al. 2010; Bourdon et al. 2010) although this may simply indicate that the heating timescales were long enough for gas-grain equilibration (see e.g. Richter 2004).

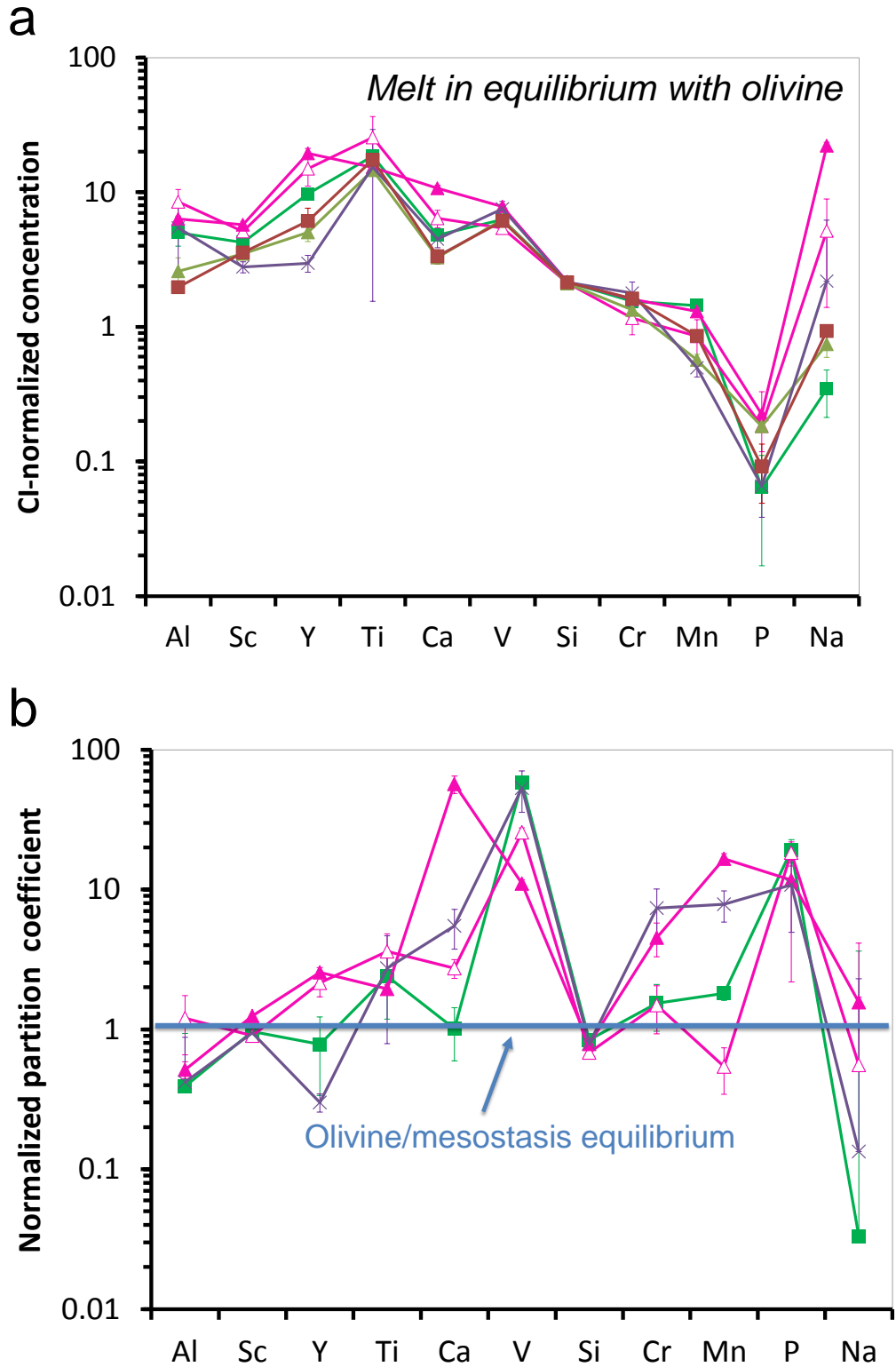

**Figure 6:** (a) Calculated composition of melt in equilibrium with olivine in the different chondrules analyzed. Partition coefficients are taken from the PO49 run of Kennedy et al. (1993) with the addition of values for P and Na from Brunet and Chazot (2001) and Mathieu et al. (2011), respectively. The melt has suprachondritic abundances of refractory elements. (b) Olivine/mesostasis partition coefficient normalized to the above experimental equilibrium partition coefficients (same

color coding as in panel a). The data generally scatter around unity diagnostic of equilibrium but show strong anomalies for several elements (see text). Error bars are one standard deviation.

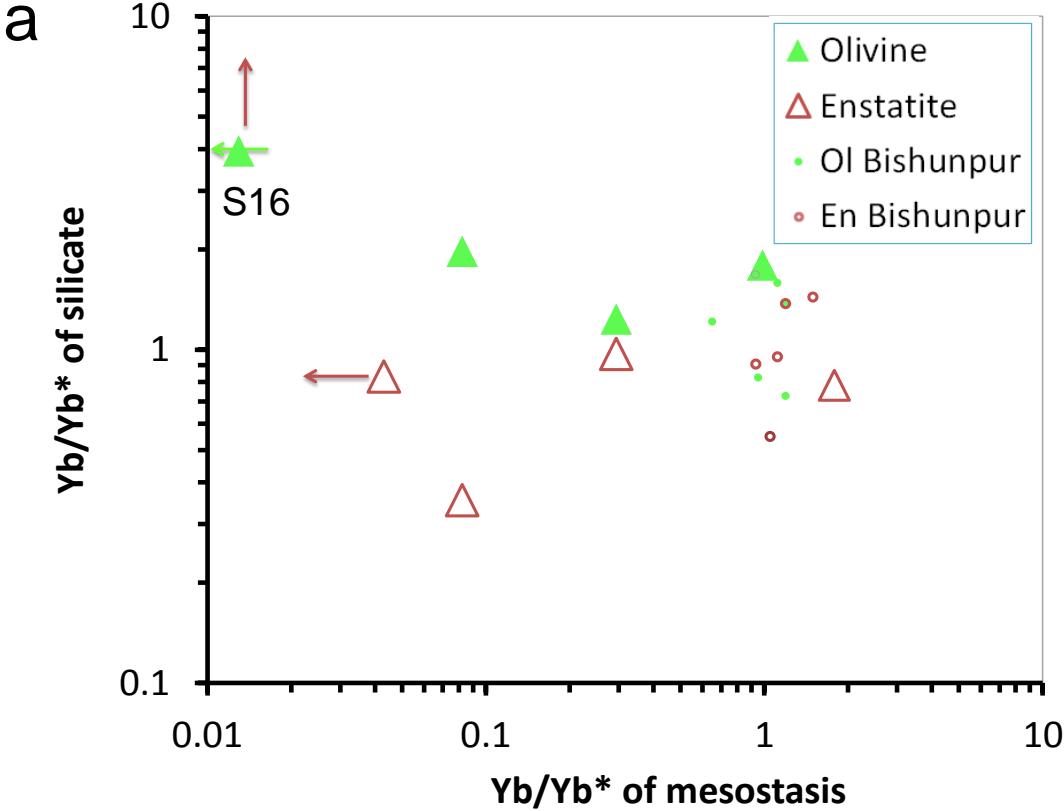

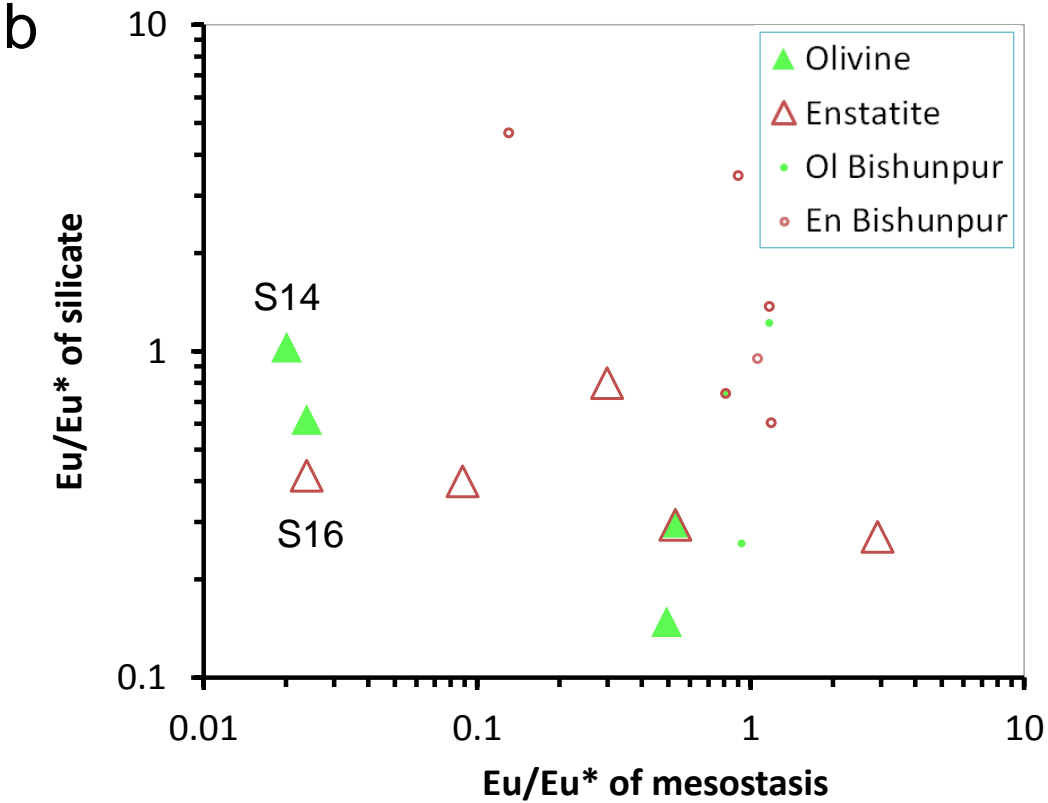

**Figure 7**: Rare earth element anomalies (a) Yb/Yb* for mesostasis vs. the same for olivine and enstatite. (b) Same for Eu/Eu*. Also plotted for comparisons are data for Bishunpur (LL3) type I chondrules (Jacquet et al. 2015). Arrows signal upper and lower limits.

## 4.2 Abnormal REE patterns

In this section we discuss the anomalous LREE-enriched patterns of olivine and enstatite in chondrules S14 and S16. So unusual a feature certainly invites extra caution to exclude any potential analytical bias, e.g. contamination. However, Ca and Al are only slightly elevated (e.g. 0.13 and 0.21 wt% Ca in S14 and S16 olivine, respectively, compared to an average 0.08 wt% for the other) and would allow very limited contributions of REE-rich carriers like mesostasis, oldhamite, or terrestrial weathering products. Independently of absolute concentrations, the very shapes of the REE patterns, especially those in S16 with their Eu and Yb anomalies, are inconsistent with any of these possibilities. Adding to this the multiplicity of analyses yielding similar patterns in each chondrule (not to mention analyses rejected because of contamination), we are confident that these REE patterns are intrinsic to the minerals in question.

The array of REE patterns in chondrule S16 (Fig. 3b) strongly suggests that the different features result from anomalous *internal* partitioning of REE between the phases: positive Sm and Yb anomalies in olivine and pyroxene mirror the negative ones in mesostasis, and likewise the LREE enrichment of the ferromagnesian silicates is complemented by an increase from LREE to HREE of the mesostasis ($(La/Lu)_N$=0.5). In fact, a reconstructed bulk composition using the observed mineral modes exhibits a flat REE pattern with only a negative Eu anomaly (Fig. 3b). Adding to that the clear igneous texture of S16, it is hence difficult to envision that chondrule S16 results from the chance, unequilibrated aggregation of solids from different stages of a noncanonical condensation sequence, say at high C/O ratios, which would hardly account for the mineralogy of S16 (devoid of carbides or oldhamite; see e.g. Ebel 2000) anyway. The REE patterns must then reflect igneous partitioning anomalies in Yb, Sm as well as Eu (since, for the latter, a positive anomaly in the olivine/mesostasis and enstatite/mesostasis partition coefficients is in order to essentially cancel the negative anomaly of the mesostasis in the resulting REE patterns for both silicates). The most natural explanation for these is that these REE were in the divalent form; and their partition coefficients would indeed roughly fit in the Onuma diagram for divalent ions (Fig. 5c, d). This is in fact what we already inferred for Eu and Yb during oldhamite formation above, and would be the most direct evidence that this valence state indeed occurred. As to Sm, to our knowledge, this would be the first unambiguous evidence for $Sm^{2+}$ in natural minerals, although such presence had been originally suggested by Goldschmidt (1954). Perhaps some of the anomalies in such elements witnessed by Hsu and Crozaz (1998) in their pyroxene analyses indicate that these were divalent also (they then deemed this unlikely given that anomalies do not occur in all enstatite grains, but we see no reason to assume a priori that all enstatite chondrite chondrules formed under the same redox conditions, or underwent the same degree of equilibration, especially given the existence of FeO-rich silicates). This would call for reducing conditions being *recorded* by REE patterns in ferromagnesian silicates in this chondrule, in contrast to all other olivine-bearing chondrules analysed here. It seems impossible at this point to be more precise on what is really implied by these "reducing conditions", for while much work has focused on the valence state of Eu as an oxybarometer (with the IW buffer roughly corresponding to the $Eu^{3+}/Eu^{2+}$ transition; see Shearer et al. 2006), we are aware of no such study for Yb (and a fortiori for Sm) in silicate materials. In fact, the enstatite/melt partitioning experiments of Cartier et al. (2014) show *no* evidence for significant Yb anomalies for oxygen fugacities down to IW-8.2 (although a Eu anomaly was always observed); either the presence of $Yb^{2+}$ requires more extreme conditions (in a way or another), or kinetic inhibitions may have been at play for the timescales of these runs. A calibration of conditions for divalent Yb would obviously be important, since we are also more generally arguing for its presence during oldhamite formation.

Such valence changes cannot however be an explanation for the *negative slope* of the REE patterns of the ferromagnesian silicates, for no Eu, Yb and/or Sm anomaly is present in S14 olivine which also shows this feature. On the Onuma diagrams for trivalent ions (Fig. 5a,b), the data for S14 and S16 seem to indicate the existence of a second crystallographic site of radius ~1 Å (in addition to the "regular" M2 one at ~0.7 Å). The trend is less well-defined for the divalent ions (Fig. 5c,d) although there certainly is a distinct enhancement of the partition coefficients as well around that size for S14 and S16 (we note that this extends to Pb (ionic radius 1.19 Å), excluded from the plot because of its anomalous behaviour (Wood and Blundy 2003), which is one order of magnitude more compatible in S14 and S16 silicates than in other chondrules (see Table 2)). We are aware of only one experimental olivine/melt partitioning study yielding comparable results, namely that performed by Saito et al. (1998) using L6 chondrite Allan Hill 76009 as a starting material. While normal HREE-enriched patterns were obtained at 1320°C, runs at 1440°C —that is near-liquidus — yielded a negative REE slope (with a upturn around Tm), with intermediate results at intermediate temperatures. Saito et al. (1998) could not narrow down the physical cause of this LREE enrichment, although they offered speculations in terms of lattice defects behaviour at high temperatures. Perhaps it could relate to incorporation of REE *during* crystallization (competing with dissolution near liquidus) as opposed to substitution in a pre-existing lattice. It remains however uncertain whether the Saito et al. (1998) results are relevant here; while chondrules S14 and S16 are the most mesostasis-rich of the series (with 31 and 16 vol% mesosatis, respectively), which could suggest efficient dissolution of precursor crystals, their textures do not seem particularly remarkable as far as porphyritic chondrules are concerned (see Fig. 1e,f).

It may be that these REE "oddities" are actually linked to a more general feature. Indeed, "normal" olivine and low-Ca pyroxene in chondrules almost always show higher LREE abundances than expected from equilibrium partitioning (Alexander 1994; Jones and Layne 1997; Ruzicka et al. 2008; Jacquet et al. 2012, 2015; this work). Kennedy et al. (1993) suggested that, at least in their experiments, these reflected glass contamination, which would account for the flattening of the REE patterns for LREE (mirroring that of the glass). However, at least for enstatite, there is a correlation between LREE and HREE (Jacquet et al. 2015) which would be difficult to understand if LREE were carried by the glass whereas the HREE budget would be dominated by bona fide pyroxene. Also, there seems to be a tendency for LREE to frequently exhibit a "spoon shape", that is a distinct decrease from La to about Pr before an upturn and an enrichment (see e.g. Fig. 9 of Jacquet et al. 2012; or the LL chondrule olivine average in the Onuma diagram on Fig. 5 herein). This feature not uncommonly appears for terrestrial experimental and natural data as well (Bédard 2005). While increasing the apparent partition coefficients, incorporation of mesostasis would not change their relative order for different elements (unless there is a relative sensitivity change for LREE analyzed by LA-ICP-MS from glass on the one hand to olivine and pyroxene on the other hand). Same would apply for the alternative possibility of a boundary layer effect (e.g. Albarède 2002). However, a second crystallographic site as suggested for chondrules S14 and S16 could explain such an inversion of the order of incompatibility. That is, it could be that the same phenomenon, whatever its exact nature, applied to most chondrules to various (generally incipient) degrees. As a circumstantial evidence, if this, again, is at all to be related to the Saito et al. (1998) experiment, we note that olivine in all 5 barred olivine chondrules with detected LREE analyzed in our past studies (Jacquet et al. 2012, 2015) have $(La/Ce)_N > 1$ (that is, a relative enrichment in the lightest REE) whereas the value of 1 is the median for the general olivine population; this could be linked to their thermal history involving efficient melting and undercooling. Whatever that may be, given the enduring uncertainty on the actual cause of LREE overabundance, we advise anew against use of LREE concentrations to calibrate cooling rates of chondrules (Jacquet et al. 2012).

## 4.3 Ferroan pyroxene

Beside olivine, ferroan pyroxene is another witness of relatively oxidizing conditions. In this section, we briefly discuss our analyses of such material.

Let us start with the pyroxene-silica clasts. The fact that their silica "worms" are often surrounded by radial cracks (see e.g. Fig. 1g) is suggestive of volume increase after retromorphosis from a denser high-pressure form of silica, e.g. stishovite or seifertite (El Goresy et al. 2008), subsequent to a shock event. We note that Kimura et al. (2005) only reported quartz, cristobalite, tridymite and glass in their survey of enstatite chondrite silica, but their study did not include Sahara 97096 nor any object comparable to our clasts. The purported shock event, whether related to the individuation of the clast or not, could explain the elevated and flat (notwithstanding possible Eu anomalies) REE pattern of pyroxene (Fig. 2c), for shock may have caused injection of incompatible element-enriched melt in it, as reported by Ruzicka et al. (2008) for olivine in the LL3.2 chondrite Sahara 97210 (shock stage S4). Indeed, Al concentrations in the pyroxene are high (0.4-1.6 wt%) and could indicate contaminations at tens of % levels. We thus conclude that the pyroxene-silica clasts were recrystallized by shock events.

We now turn to the ferroan pyroxene spherules (Fig. 1h). It is tempting to attribute their negatively sloped REE pattern to a similar process as the previous section, be it near-liquidus partitioning or something else, with the proviso that the present-day spherules only sample the pyroxene part of a larger igneous system. Nonetheless, the positive Ce and negative Eu anomalies of S25 would point to vapour/condensed phase processes, as Ce becomes more volatile and Eu more refractory with increasing oxygen fugacity (and vice-versa; Boynton 1989), which processes may have pre- or post-dated igneous partitioning. It is also conceivable that the spherules originally formed as condensates, so long incorporation (and therefore fractionation) of REE was controlled by the pyroxene structure, yielding thus similar fractionations than partitioning with a melt, on top of "pure" volatility trends (see Boynton 1989). Whatever that may be, the spherules may be genetically linked to the pyroxene-silica clasts (which have similar ferrosilite contents) and could possibly be microtektites or microkrystites originating from closer to the purported impact points, perhaps owing their ferrosilite contents to oxidizing conditions in the impact plume. But such thoughts are only speculations at this point. We are content to recall that oxygen isotope compositions of these pyroxenes are indistinguishable from that of enstatite (Kimura et al. 2003), indicating formation from local material, and that shock seems to have been quite prevalent on enstatite chondrite parent bodies compared to other chondrite groups (Rubin et al. 1997).

## *5. Conclusion*

We have performed in situ LA-ICP-MS analyses of olivine, pyroxene and mesostasis in chondrules in the EH3 enstatite chondrite Sahara 97096, focusing on olivine-bearing chondrules. Our main observations are the following:

1. Most olivine and enstatite have REE patterns comparable to their counterparts in type I chondrules from other chondrite groups (e.g. ordinary chondrites)
2. Mesostasis frequently exhibit negative anomalies in Eu, Yb and/or Sm.
3. In two chondrules, olivine and enstatite have REE patterns decreasing from LREE to HREE. In one of these two (S16), they exhibit positive anomalies in Sm, Yb and likely Eu complementary to negative anomalies of the mesostasis.
4. We also analyzed some ferroan pyroxene inclusions. Pyroxene shows flat REE patterns near chondritic levels for pyroxene-silica clasts and negatively sloped REE patterns for spherules.

We offered the following interpretations:

1. Olivine and enstatite mostly record igneous formation under redox conditions similar to those of type I chondrules in other chondrites.
2. Mesostasis evolved out of equilibrium with the above, possibly by partitioning with oldhamite during a sulfidation event as inferred by Lehner et al. (2013). Anomalies in Eu, Yb and/or Sm possibly arose because these REE were in the divalent state. Oldhamite was likely expelled from the chondrules hence the abundance of opaque nodules outside them in EH chondrites.
3. The exceptional negatively sloped REE patterns may be related to the frequent (albeit more limited) LREE enrichment of chondrule ferromagnesian silicates relative to lattice strain model predictions, possibly as a result of near-liquidus partitioning.
4. Ferroan pyroxene in Sahara 97096, especially the pyroxene-silica clasts, may have originated by impact.

We thus support the emerging picture that a condensation sequence at elevated C/O ratios may not be required to account for the unique characteristics of enstatite chondrites, and that they may have been established by exposure of "normal" (chondrite-wise) material to an extreme sulfidizing environment during chondrule/opaque nodule formation. Its actual physico-chemical conditions have however yet to be precisely determined.

*Acknowledgments*: Reviews by Stephen Lehner, Michael Weisberg, an anonymous referee and the AE Christine Floss were greatly appreciated, in particular to improve the discussion of partition coefficients and the properties of opaque assemblages. We thank Laurette Piani for fruitful discussions on Sahara 97096 and the genesis of enstatite chondrites in general. We are also grateful to the Muséum National d'Histoire Naturelle for the sample loan.